\documentstyle[aps,epsf,twocolumn,prl]{revtex}                                         

\begin{document}
\draft

\twocolumn[\hsize\textwidth\columnwidth\hsize\csname@twocolumnfalse\endcsname

\title
{Non-Fermi-Liquid Scaling in 
Ce(Ru$_{0.5}$Rh$_{0.5}$)$_2$Si$_2$}

\author{Y. Tabata$^1$, D. R. Grempel$^{2,}$\cite{add1}, M. Ocio $^{2}$, T. Taniguchi$^1$
and Y. 
Miyako $^1$}

\address{
$^1$Graduate School of Science, Osaka University, Toyonaka, Osaka 560, Japan\\
$^2$CEA/Saclay, Service de Physique de l'Etat Condens\'e, 91191 
Gif-sur-Yvette Cedex, France 
}

\date{\today}
\maketitle 
\widetext 
\begin{abstract} 
\noindent 
We study the temperature and field dependence of the
magnetic and transport properties of the non-Fermi-liquid compound
Ce(Ru$_{0.5}$Rh$_{0.5}$)$_2$Si$_2$. For fields $H \lesssim $ 0.1 T the results
suggest that the observed NFL behavior is disorder-driven. For higher
fields, however, magnetic and transport properties are dominated by 
the coupling of the conduction electrons to
critical spin-fluctuations. The temperature dependence of the susceptibility as
well as the scaling properties of the magnetoresistance 
 are in very good agreement with the predictions of recent dynamical mean-field theories
of Kondo alloys close to a spin-glass quantum critical point.
\end{abstract}

]

\narrowtext

The properties of a large class of {\it f}-electrons materials show
striking departures from the predictions of standard Fermi-liquid
theory  at low temperature \cite{maple}. Several mechanisms leading to non-Fermi-liquid
(NFL) behavior have 
been proposed. In systems close to a quantum phase transition such as
CeCu$_{6-x}$Au$_{x}$\cite{lohn} or CeIn$_{3}$\cite{mathur}, NFL behavior is due to the coupling of the
conduction 
electrons to critical spin fluctuations \cite{millis}. Anomalous  
properties are observed when the system is driven through the quantum
critical point (QCP) by alloying
or by applying pressure. In other systems, such as
UCu$_{5-x}$Pd$_{x}$ \cite{vollmer}, NFL properties are thought to be
a consequence of the interplay of strong structural disorder and many body
effects \cite{bernal,castro}. 
In this paper we report results of a study of the temperature and field dependence  of the magnetic and
transport properties of
Ce(Ru$_{0.5}$Rh$_{0.5}$)$_2$Si$_2$. We found that this system exhibits
different types of anomalies depending on the value of the applied
field. At weak fields, we found signatures of disorder
effects such as a diverging low-temperature susceptibility and an
anomaly in the low-field 
magnetoresistance. Above
1kG, the $T$ and $H$-dependence of the
susceptibility and the resistivity agrees with the  predictions of recent 
mean-field theories of the spin-glass (SG) QCP
\cite{sachdev,georges,grempel}. The magnetoresistance is found to exhibit universal
scaling properties as predicted by the
theory.

In the Ce(Ru$_{1-x}$Rh$_x$)$_2$Si$_2$ alloy series the 
Ce-sublattice is preserved and the hybridization between 4{\it f} 
and conduction electrons varies with the concentration of the 
ligand 4{\it d} atoms 
Ru 
and Rh. Pure 
CeRu$_2$Si$_2$ is a heavy fermion compound with a  $\gamma$ value of
about 385 mJ/mol/K$^2$ \cite{besnus} and no long range magnetic order down to 
20 mK.  With substitution of Rh for  
Ru, a spin density 
wave (SDW) region appears between $x=0.03$ and $x=0.4$  
\cite{miyako1,miyako2}.  While Ce(Ru$_{1-x}$Rh$_x$)$_2$Si$_2$ at $x =
0.03$ is a normal Fermi liquid
\cite{tabata,take}, a NFL regime
exists for $x = 0.4$ and 0.5 \cite{take,taniguchi}. In pure CeRh$_2$Si$_2$, 
the 4{\it f} electrons are localized and the material is antiferromagnetic (AF) 
below $T_N = 35$ K 
\cite{quezel}. With increasing Ru substitution, $T_N$ decreases and
eventually vanishes at a critical 
concentration $x_c \approx 0.5$ \cite{kawa}. 
There is no direct evidence of long range AF order 
below $x = 0.7$. The admixture of
Ru and Rh introduces magnetic frustration effects as it leads to
a competition 
between widely different
types of magnetic short-range  
order \cite{taniguchi}. A frustrated ground state of the SG type can
not be excluded
 slightly above $x_c$. Recent $\mu$-SR studies \cite{yamamoto} showed that the
$T$-dependence of the muon relaxation rate in the $x=0.5$ alloy is similar to that
observed in spin glasses. The muon depolarization rate decays
exponentially, however, showing that the spin correlations are dynamic
rather than static.

Samples of Ce(Ru$_{0.5}$Rh$_{0.5}$)$_2$Si$_2$ were prepared by arc-melting of the constituents in an 
argon atmosphere. The ingots were remelted several times to insure homogeneity. 
Single 
crystals oriented 
along {\it a} and {\it c}-axis were grown 
by the Czochralsky method in a tri-arc furnace in argon atmosphere and
parallelepiped-shaped 
samples of size 
$\approx 0.5\times 
0.5\times 4$ mm$^3$ were obtained. 
The resistivity was measured by a 
standard ac method 
at 17 Hz in the range 16 mK - 4 K and in magnetic fields up to 5 T. 
The low-field susceptibility was measured in the dilution range 
$\approx$ 50 mK - 4 K using a 
standard 
method at 130 Hz. The static magnetization was 
measured in the 
same range of temperature with a SQUID magnetometer in a dilution
setup and,  
above 3 K, in a commercial SQUID magnetometer.

\begin{figure}
\epsfxsize=3.5in
\epsffile{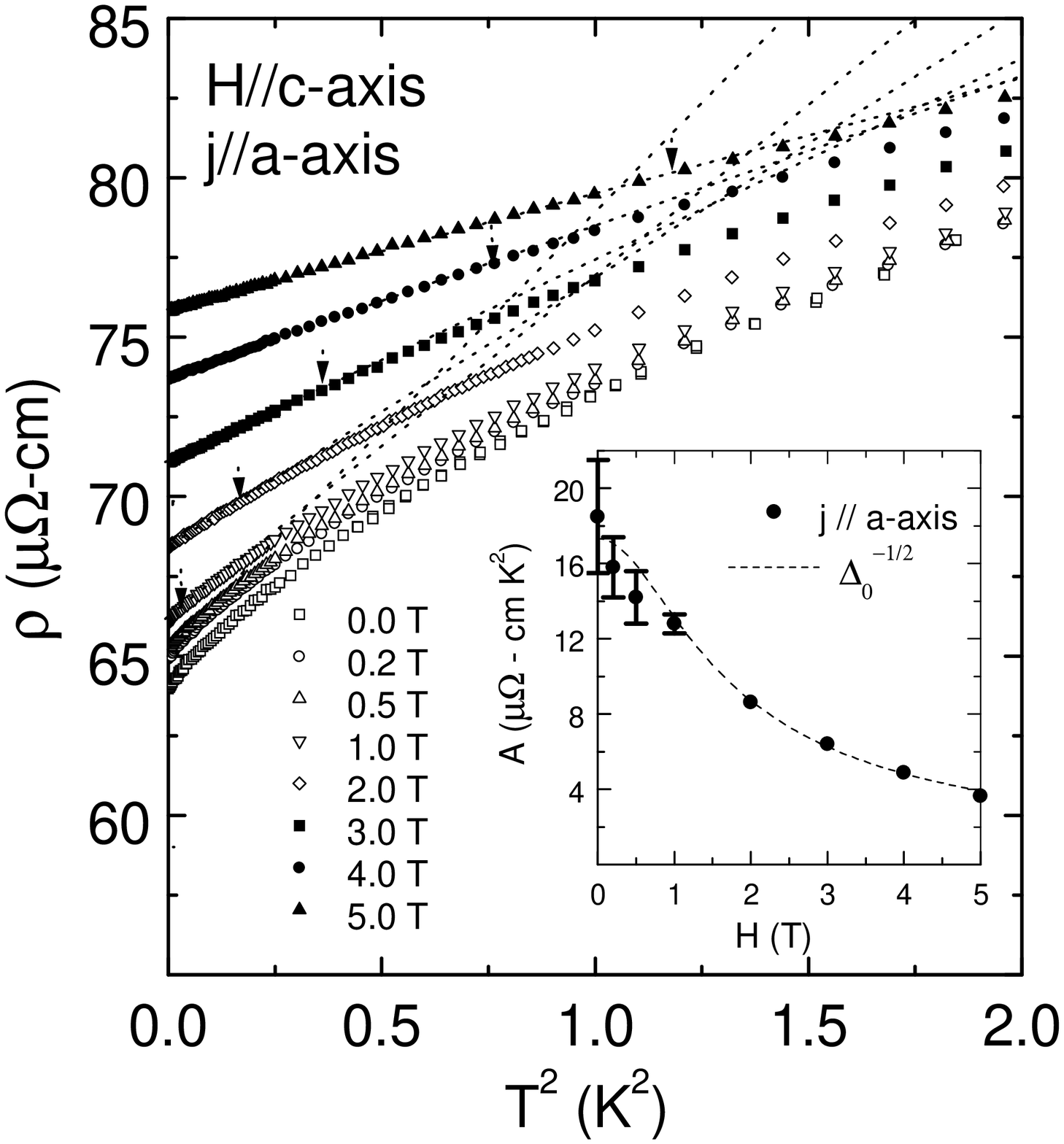}
\caption{Resistivity of Ce(Ru$_{0.5}$Rh$_{0.5}$)$_2$Si$_2$ in a
 magnetic field as a function of $T^2$. The magnetic field is  along
the {\it c}-axis and the current flows along the {\it a}-axis. The arrows indicate the range of temperature
where the resistivity varies as T$^2$. Inset: field-dependence of
$A\equiv d\rho/dT^2$ determined from the data below 100 mK.}
\label{rho_T2}
\end{figure}

Fig. \ref{rho_T2} displays the resistivity $\rho$ measured in a field
applied along  
the {\it c}-direction up to 5 T. The data correspond to a current flow
in the {\it a}-direction. The behavior
along the $c$-axis is similar  
but the resistivity is about four times
smaller. The 
magnetoresistance is positive at low temperatures and changes sign at 
about 2.5 K 
in the {\it a}-direction and at about 1.5 K in the {\it c}-direction. 
In both cases the data at 5 T follow a $T^2$-law up to T $\approx$
1 K. The range of temperatures in which FL 
behavior is observed decreases with field and 
vanishes at $H=0$. 
These results are qualitatively similar to those obtained in CeCu$_{6-x}$Au$_{x}$\cite{lohn}
and in the stoichiometric compound 
CeNi$_2$Ge$_2$  \cite{steglich}.  The resistivity of our samples is
much higher, however, as a consequence of the high degree of  
substitutional disorder present in the $x=0.5$ alloy.

The temperature dependence of the zero-field resistivity is $\delta
\rho \propto T^{1.6}$ as shown in  
Fig. \ref{magrho}(a). The resistivity exponent is close but not
identical (see below) to that expected for metallic antiferromagnets \cite{millis} and
spin-glasses \cite{sachdev,georges} at the QCP in the high resistivity
limit \cite{rosch}.
The low-temperature magnetoresistance $(\rho (H)-\rho (0))/\rho (0)$ along the
$a$-direction is plotted in Fig. \ref{magrho}(b) for $H < 1$ T. At
high fields (not shown in the figure) it
shows the  classical $H^2$-dependence due to the bending of the
electron orbits. Below a few kG, however, the low-temperature resistance increases
{\it linearly} with field as $H\to 0$. 

\begin{figure}
\epsfxsize=3.2in
\epsffile{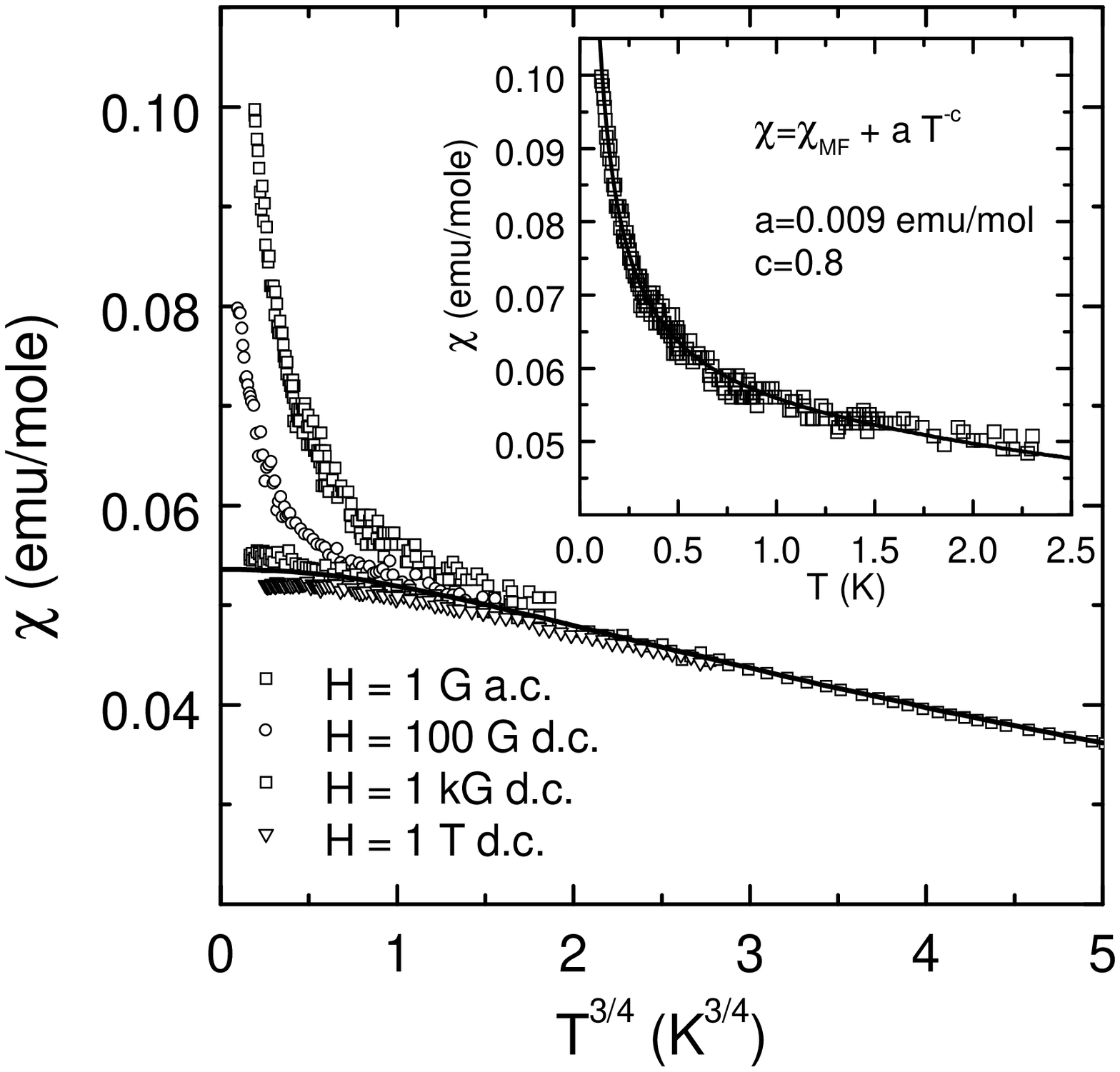}
\caption{Susceptibility M/H measured in several magnetic fields along
the {\it c} direction, plotted as
a function of T$^{3/4}$. The solid curve is a fit of the data at $H$=1
kG to the expressions in  Eqs. (\protect{\ref{delta}}) and (\protect{\ref{chistatic}}). Inset: The ac susceptibility at $H$=1
G. The solid line is the fit mentioned in the text.}
\label{suscept}
\end{figure}

The susceptibility $\chi$ = M/H (M is the magnetization) is 
represented in Fig. \ref{suscept} for H$_{ac}$ = 1 G
and H$_{dc}$ = 0.01 T, 0.1 T, and 1 T for $T <$ 10 K. 
At 
low temperatures 
$\chi$ decreases strongly with increasing field between 1 G and 1 kG 
but weakly above 1 kG.  Above 3 K the field has no
effect 
on $\chi$ up to 1 T as seen by comparison of the curves at 1 kG and 1 T in
Fig.\ (\ref{suscept}).
At 1 G, $\chi$ increases sharply with decreasing $T$ below 2 K. 

The divergence of $\chi$ at low-fields as well as that of $C/T$
\cite{taniguchi} 
suggests that, in this regime, NFL
behavior may be driven by disorder \cite{bernal,castro}. 
The linear 
(rather than quadratic) rise of the low-field 
magnetoresistance at low
temperatures reported here may be explained by 
Kondo-disorder effects \cite{ohkawa}. 
Recent $\mu$SR experiments \cite{yamamoto2}
showed a sharp increase in the muon relaxation rate below 2
K that saturates at a $T$-independent value below 0.7 K. 
This behavior has been interpreted in terms of the Griffiths-phase
scenario \cite{castro} in which finite clusters carrying a magnetic
moment 
fluctuate very slowly at low
temperature due to quantum tunneling. Magnetic, NMR and
specific heat experiments performed in a temperautre range much higher
than ours have also been interpreted in terms of this mechanism \cite{graf,mac}.

\begin{figure}
\epsfxsize=3.5in
\epsffile{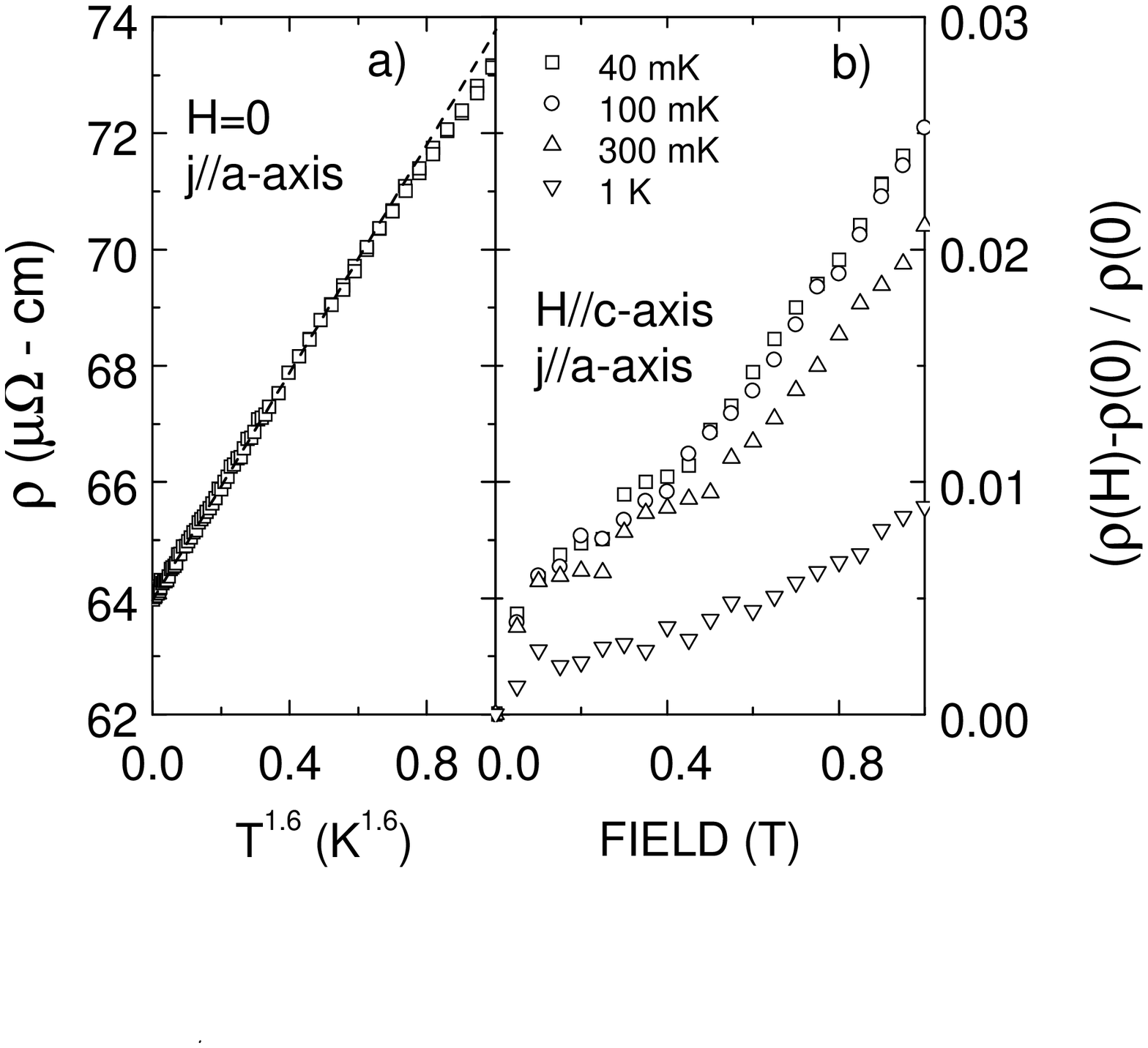}
\caption{(a) Temperature dependence of the zero-field resistivity of
Ce(Ru$_{0.5}$Rh$_{0.5}$)$_2$Si$_2$ along the $a$-direction.(b) Magnetoresistance $\rho (H)-\rho (0))/\rho (0)$ measured along the 
{\it a}-direction at several temperatures and 
plotted as a function of the field applied in the {\it c}-direction.}
\label{magrho}
\end{figure}

For $H \gtrsim$ 1 kG, $\chi$ remains finite as $T\to 0$ and depends
weakly on $H$, suggesting that application of a
moderate magnetic field quenches the mechanism that leads to the
divergence observed at lower fields.
Although a full understanding of this fact is still
lacking, it should be noticed that in the Griffiths-phase
model \cite{castro} quantum fluctuations of the largest clusters 
are expected to be suppressed by a small magnetic field.
Indeed, while the Zeeman energy of a cluster of size $N$ grows as
$\sqrt{N}$, its tunneling energy vanishes exponentially with $N$.
In the following we concentrate on the physics above 1\ kG.

The $T$-dependence of the susceptibility at 1 kG is
still anomalous and $\delta \chi(T)$ is approximately linear
in $T^{3/4}$
(cf. Fig. \ref{suscept}) except at the lowest temperatures, a point
that will be 
discussed further below~\cite{note}. A $T^{3/4}$ dependence of $\chi$ as well as
the 
value 3/2 for the resistivity exponent where predicted by recent dynamical mean
field-theories (DMFT) of the spin-glass QCP
\cite{sachdev,georges,grempel}. In view of the important role 
that frustration is expected to play
 in this compound it is tempting to try to interpret our results
in terms of the fully frustrated SG model. The latter describes
 conduction band electrons coupled to localized $f$-electron spins
 {\it via} a local Kondo coupling, $J_{\rm K}$. There is, in addition, a
 residual Ising-like interaction between the localized spins. The
 effects of the magnetic frustration introduced by disorder 
 are incorporated by taking 
 random spin couplings $J_{ij}$ chosen from a symmetric distribution of
 width $\left< J_{ij}^2 \right> = J^2$. Disorder in the Kondo
temperature is not included in the model. From the Kondo-temperature
distribution determined in
Refs. \cite{graf} and \cite{mac} one can conclude that 
this effect should play a lesser role than frustration in the
low-temperature range 
that interests us.
The SG model \cite{georges,grempel} was  investigated in 
the framework of dynamical mean-field theory \cite{rmp}. It exhibits a
zero-temperature SG transition when the typical exchange energy $J=J_{\rm c} \sim
T_{\rm K}$, the Kondo temperature of the underlying
Kondo lattice. Monte Carlo
 simulations \cite{grempel} of this model showed that its critical 
properties are described by an effective strong-coupling 
theory closely related to other mean-field 
models\cite{sachdev,georges}. The physical
 properties of the system depend on the effective distance to the
QCP, $\Delta(T,H)$. It can be shown that this is 

\begin{equation}
\label{delta}
\Delta = \Delta_0 + 2 \sqrt{\Delta_0} \frac{T}{T_0} \left[\sqrt{1+\frac{ 
T}{2\sqrt{3}\,  
T_0\Delta_0}} - 1\right],
\end{equation}
where $\Delta_0= (1-J/J_{\rm c})+ (H/H_0)^2 $ and $H_0 = J_{\rm c}/(g \mu_{\rm
B})$ ($g$ is the gyromagnetic ratio of the Ce ion). The 
scale $T_0$ is proportional to $T_K$. Numerical
simulations yield $J_{\rm c}\sim 1.15\  T_{\rm K}$. The spin 
susceptibility is \cite{grempel}

\begin{equation}
\label{chistatic}
J_{\rm c}\chi=\sqrt{1+\Delta} - \sqrt{\Delta}.
\end{equation}
The spectrum of magnetic excitation has a scaling form
\cite{sachdev,georges,grempel}, 
$J_{c} \chi''(\omega)=\sqrt{\Delta} \Phi(\omega/J_{c}\Delta)$,
where the  universal scaling function $\Phi(x)=x/\sqrt{2}\ [
(1+x^{2})^{1/2} + 1 ]^{-1/2}$.
The temperature-dependent contribution to the resistivity 
is computed from  
$\delta\rho \propto 1/\tau$ with the inverse scattering time
$\tau^{-1}\propto \int_0^{\infty} d\omega
\chi''(\omega) /\sinh(\beta\omega)$, an expression valid in the dirty
limit \cite{rosch}. The resistivity has the scaling form 
\begin{equation}
\rho(T,H) - \rho(0,H) \propto T^{3/2}\Psi\left(\frac{T}{\Delta T_0}\right),
\label{scaling}
\end{equation}
with $\Psi(x)=x^{-1/2}
\int_0^{\infty} du  \Phi(u x) /\sinh u$. It follows that
 $\delta\chi\equiv\chi(0)-\chi(T)\propto T^{3/4}$ and $\delta \rho \propto T^{3/2}$ at
the QCP. The resistivity
exponent of the
mean-field SG model coincides with that of the $d=3$, $z=2$
antiferromagnet \cite{sachdev,georges}. The susceptibility exponent is
specific to the SG model. Away from the QCP, normal 
Fermi-liquid behavior as $T\to 0$ is recovered with both 
$\delta \chi$ and $\delta \rho  \propto T^{2}/\sqrt{\Delta_0}$ at
sufficiently low $T$.  The crossover between these limiting
forms will be discussed below. The parameters of the theory 
can be determined from an analysis of the experimental data. 
At the critical concentration, $r=1-J/J_{\rm c}$
vanishes. However, $x_{\rm c}$ is not known accurately and a small but 
finite $r$ can not be excluded {\it a priori}. The
characteristic field $H_0$ may be estimated 
from an extrapolation to $T=0$ of the susceptibility per Ce atom. From the definitions 
above and Eq.\ (\ref{chistatic}) we estimate $H_0=\mu_{\rm B}/\chi(0)\approx
11$ T (we have assumed $g=2$). Since $H_0$ is very
 large, the measuring field can be neglected in the analysis of the
 magnetization at 1 kG. A fit of $\chi$ using Eqs.\
(\ref{chistatic}) and \ (\ref{delta}) with $\Delta_0=0$ gives 
$T_0 \approx 20 $K, which is slightly smaller than the Kondo temperature of the system estimated from the $T$-dependence of the resistivity\cite{take}. 

\begin{figure}[htb]
\epsfxsize=\hsize
\epsfbox{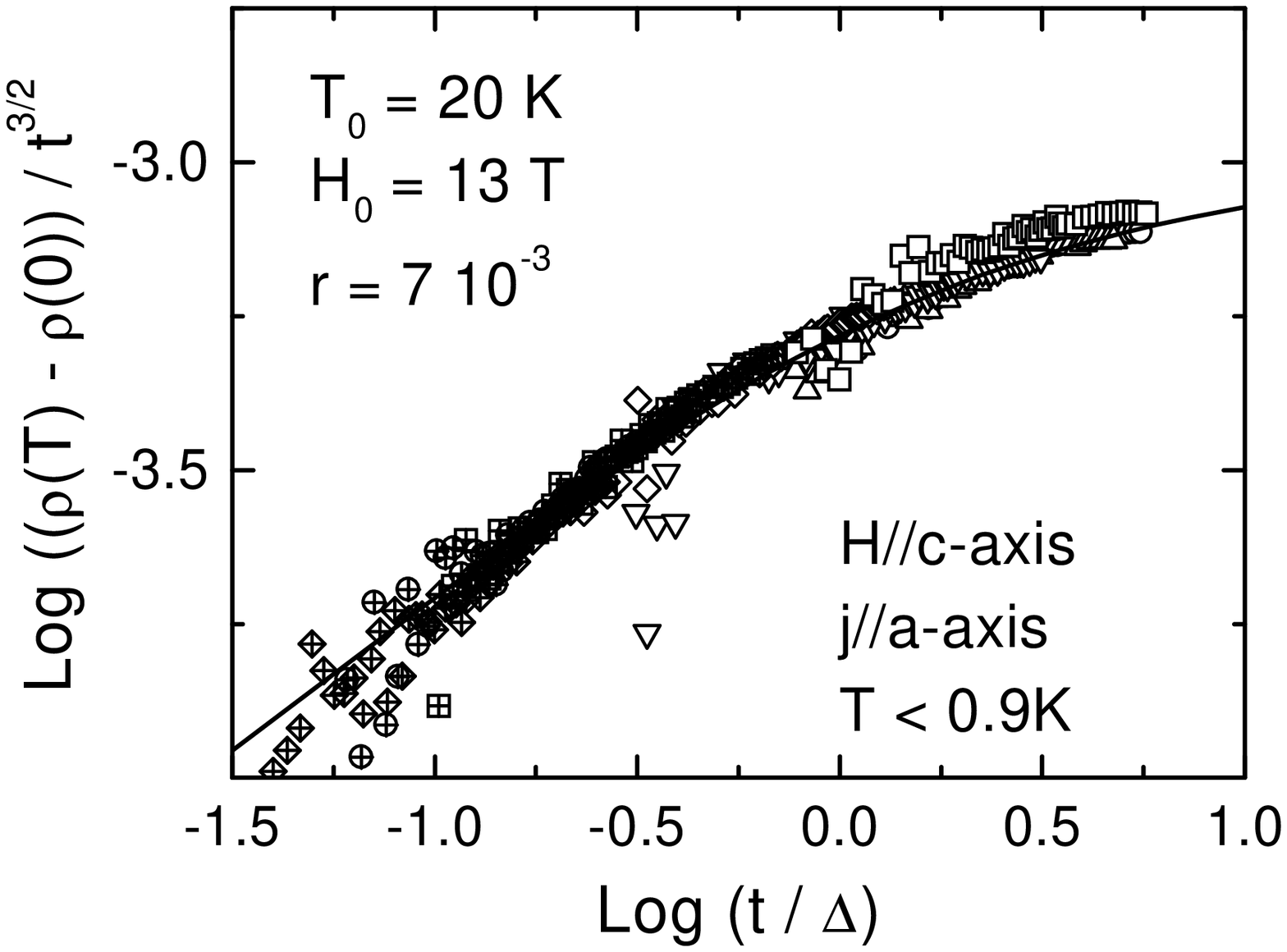}
\caption{Scaling plots of the resistivity along the $a$-axis. The
solid line is the theoretical scaling curve.}
\label{scalr}
\end{figure}

We analyzed the $T$- and $H$-dependence of the
resistivity using Eqs.\ (\ref{scaling}) and
(\ref{delta}). The condition
that all the data in Fig.\ \ref{rho_T2} collapse into a single
scaling curve leaves little freedom in the choice of the
parameters. In particular, we found it impossible to scale all the
curves with $r=0$.
A scaling plot of the resistivity 
along the $a$-axis is shown in Fig.\ \ref{scalr}. The data points are the values  of the scaled 
resistance $(\rho(T,H)-\rho(0,H))\times T^{-3/2}$ plotted {\it vs} the reduced 
variable $t/\Delta$ ($t=T/T_0$) for $T\le 0.9$
K and $H\le 5$ T. The values of the parameters are $r=7\times
10^{-3}$, $T_0=20$ K and $H_0=13$ T. The value of $r$ 
measures the distance to the true QCP, $r = \delta J/J_{\rm c}$ giving $\delta J \approx 0.2$ K, a very small
energy compared to the other energy scales present in the problem. The
characteristic field determined from this analysis is close to the
theoretical estimate given above. The solid line in Fig.\
\ref{scalr} is the theoretical scaling function $\Psi(x)$. There are 
no adjustable parameters other than an amplitude that fixes the vertical scale.
The agreement between 
theory and experiment is very good except for the data for $H$=0 
(the empty squares in Fig.\ \ref{scalr}) which lie slightly
above the scaling curve. The slope of the curve, that measures the effective
resistivity exponent, is correctly reproduced by the theory. We can
now understand that the
deviation of the resistivity exponent  (cf. Fig.\ \ref{magrho}a) with
respect to its value at the QCP (1.5) is due the small
but finite value of $r$. The effective exponent only reaches 1.5 for $t/\Delta\to \infty$, {\it i.e.}
for $r$=0.  We ascribe the excess amplitude for $H=0$, represented by the
vertical shift, to additional scattering processes dues to
disorder. We can 
compare $A$, the amplitude of the $T^2$ term in the resistivity in the FL
region, with the theoretical prediction, $A\propto
1/\sqrt{\Delta_0}$. The inset in Fig.\ \ref{rho_T2} shows the field
dependence of $A$ as determined from the initial slope $d\rho/dT^2$ of
the resistivity and the theoretical prediction. There is good agreement. 

The susceptibility can also be calculated and compared with the data. 
The solid line in Fig.\
(\ref{suscept}) is the theoretical result for $H=1$ kG.
 The data (and the theoretical curve) deviate from a pure $T^{3/4}$ law as $T\to 0$. This is due to the finite
value of $r$ which results in normal FL behavior below a crossover
temperature
 $T_{\rm FL} \sim T_0\
\Delta_0$. $T_{\rm FL}$ increases with field and can be estimated as $\approx$ 0.25 K
for $H$=1\ T.  The crossover to $T^2$ behavior in $\chi(H =1 {\rm T})$ can be 
seen in  Fig.\ (\ref{suscept}). 
The
low-field results can be described by adding to the dynamical
mean-field result an additional diverging contribution. The presence of a 
paramagnetic phase giving rise to a $T^{-1}$ divergence of $\chi$ 
was suggested in Ref.~\cite{mac}. However, to suppress such a contribution at 3 K by
a field as small as 1 kG one would need that the impurities carry
huge moments ($>$ 3 $\mu_B$). Furthermore, this hypothesis would not
explain the divergence of $\gamma$~\cite{taniguchi}.
 The inset in Fig.\ \ref{suscept} shows
a fit of the ac data for $H$=1 G to the expression
$\chi(T)=\chi_{\rm MF}(T)+ a T^{-0.8}$. This can be interpreted in
terms of the  Griffiths-phase model
\cite{castro} that predicts a power-law divergence 
$\delta \chi \propto T^{-1 + \lambda}$ with an exponent $\lambda$ that
vanishes at the QCP. The value $\lambda=0.2$ that comes out
of our analysis is consistent with this
picture for a system close to the QCP. We also found that the
$\gamma$ data~\cite{taniguchi} can be accurately 
described by
the analogous expression $\gamma(T)=\gamma_{\rm MF}(T)+ a' T^{-0.8}$ where 
$\gamma_{\rm MF}(T)\propto 1-b \sqrt{T}$ is the 
DMFT prediction\cite{georges} for $C/T$. The equality of the exponents
describing the divergent parts of $\chi$ and $C/T$ is one of the
predictions of the Grifith's phase model \cite{castro}.

In summary, we have shown that the NFL 
properties of Ce(Ru$_{0.5}$Rh$_{0.5}$)$_2$Si$_2$ are determined by disorder and 
proximity to a QCP. The effects of the two mechanisms can be
disentangled by applying a small magnetic field. At low fields, 
disorder effects 
dominate. Above 1kG, however, the temperature and field dependence of the
susceptibility are well described by the dynamical mean-field theory of
the spin 
glass QCP.  The $T$- and $H$-dependent resistivity is found to obey
 a universal scaling law. 

This work was supported in part by the Grant-in-Aid for Scientific 
Research (B) and the Monbusho International Scientific Research
Program. One of us (DRG) thanks the Newton Institute for its kind 
hospitality and A. H. Castro Neto for useful correspondence.


\begin{references}
\bibitem{note} The $T^{1/3}$-law previously
proposed~\protect{\cite{taniguchi}} only fits the data in a limited
temperature range. In addition, it lacks theoretical foundation.
\end{references}
\end{document}